\documentclass[aip, reprint]{revtex4-1}

\usepackage{graphicx}
\usepackage{dcolumn}
\usepackage{bm}

\usepackage[utf8]{inputenc}
\usepackage[T1]{fontenc}
\usepackage{mathptmx}
\usepackage{etoolbox}
\usepackage{gensymb}
\usepackage{xcolor}
\usepackage[normalem]{ulem}
\usepackage{textcomp}

\def\perkg{\mathrm{kg^{-1}}}
\def\permm{\mathrm{m^{-2}}}

\def\2D{\mathrm{2D}}
\def\3D{\mathrm{3D}}
\def\mpersec{\mathrm{m~s^{-1}}}
\def\Wpermm{\mathrm{W~m^{-2}}}

\def\VAC{V_{\mathrm{AC}}}
\def\nuAC{\nu_{\mathrm{AC}}}
\def\Vlens{V_\mathrm{lens}}
\def\Vlaunch{V_\mathrm{launch}}

\def\N2{$\mathrm{N}_2$}

\def\h2o{$\mathrm{H}_2\mathrm{O}$}
\def\co2{$\mathrm{CO}_2$}
\def\o2{$\mathrm{O}_2$}

\def\Tcom{T_\mathrm{COM}}

\begin{document}


\title{\begin{center}Focused deposition of levitated nanoscale Au droplets
    \end{center}} 



\author{Joyce E. Coppock}
 \email{jec@umd.edu}
 \affiliation{University of Maryland, College Park, MD, 20742, USA}
 \affiliation{ Laboratory for Physical Sciences, 8050 Greenmead Dr., College Park, MD, 20740, USA}
\author{B. E. Kane}
 \affiliation{Joint Quantum Institute, University of Maryland, College Park, MD, 20742, USA}
 \affiliation{ Laboratory for Physical Sciences, 8050 Greenmead Dr., College Park, MD, 20740, USA}



\begin{abstract}
We describe a method for depositing nanoscale liquid Au droplets, initially levitated in an ion trap in high vacuum, onto a remote substrate. A levitated Au nanosphere is melted, expelled from the trap, and maintained in the molten state, with a laser directed along the droplet trajectory, until it reaches the substrate and rapidly solidifies. During transit, the charged droplets are focused to a small region of the substrate with an electrostatic lens.  After deposition, the substrate is removed from the vacuum chamber and imaged and analyzed by techniques such as electron microscopy and energy dispersive spectroscopy.  Over 90$\%$ of launched particles are deposited on the substrate, and when the lens is focused, particles land in a region of diameter 120 $\mu$m after traversing a distance of 236 mm.  Our technique is of value for analysis of materials prepared or processed while levitated that can be melted. Also, Au droplets may be useful as tracers for future experiments involving smaller projectiles or oriented solids.
\end{abstract}


\maketitle 


\begin{figure*}
    \centering
    \includegraphics[width=1\linewidth]{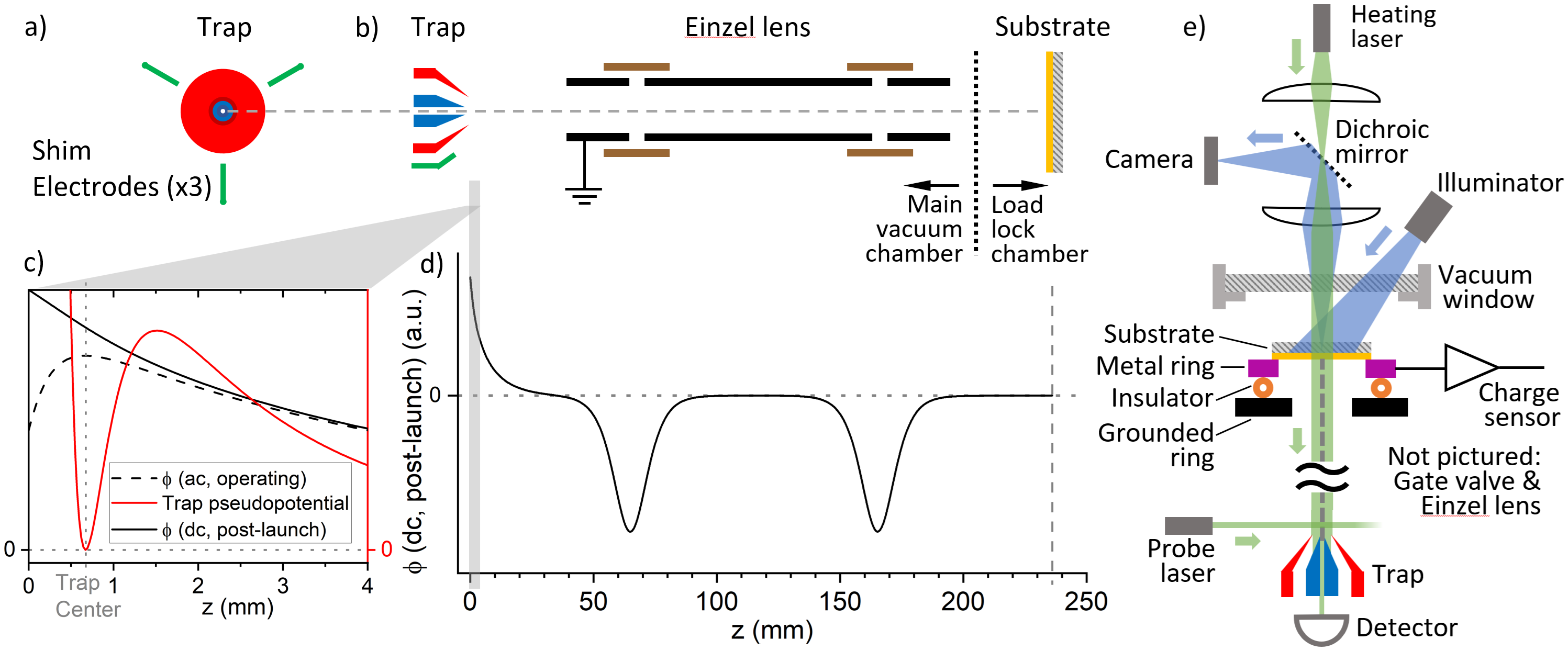}
    \caption{(a) Top view of the coaxial conical trap electrodes with three ``shim electrodes'' placed around the trap, spaced by 120\degree.  (b) Cross-sectional view of the trap, einzel lens, and deposition substrate. In the actual apparatus, the axis of symmetry (marked by dotted line) is oriented vertically.  (c) Magnified plot of potentials along the axis of the trap's conical electrodes in the vicinity of the trap.  z=0 occurs at the truncated tip of the outer conical electrode of the trap.  (d) Plot of electric potential $\phi$ along the axis of symmetry at the moment of launch.  The region highlighted in gray is magnified in part (c).  (e) Schematic of the optics and charge-sensing apparatus.  Colored arrows show direction of light paths.  The heating laser beam propagates vertically along the entire travel path of the particle and through a hole in the center of the trap.  The deposition substrate and charge sensing device are positioned inside the vacuum chamber under a vacuum window.  An illuminator ($\lambda=463 \pm 25~$nm) is directed at the substrate.  A dichroic mirror allows a camera to image the substrate along the same vertical path as the laser.  Vertical dotted line denotes the axis of symmetry of the einzel lens (see part (b)) and coincides approximately with the path of the launched particle.  All characterization prior to launch is performed with the horizontal probe laser. 
}
    \label{Apparatus}
\end{figure*}

\label{sec:intro}
Recent advances in levitation technology have opened up new  opportunities for probing nanoscale materials. \emph{In situ} measurements of levitated particle mass, as well as optical scattering, emission, and absorption,  have been used to probe surface chemistry of levitated materials, \cite{Ricci2022,Hoffman2020,Hoffman2023} nanoparticle growth and oxidation,\cite{Lau2023,Rodriguez2022} and nanoparticle thermal properties.\cite{Coppock2022}  
However, almost all probes of nanoscale materials, developed and refined over decades, can  only be used when the materials are situated on large substrates.  A method for easily depositing levitated materials onto a substrate would thus be highly desirable.

The bulk of previous work on levitated particle deposition has centered on single atoms originally confined in an ion trap.\cite{Schnitzler2010, Jacob2016}  These techniques may have applications ranging from ultramicroscopy \cite{Jacob2016} to deterministic semiconductor doping. \cite{Jacob2021}  While deposition of levitated nanoscale objects has been previously reported, \cite{Gregor2009, Kuhlicke2015,Coppock2018} a requirement for reliable deposition is that there must be sufficient adhesion between the substrate and the impinging particle to prevent the particle from bouncing off the surface.  Irregularly shaped solid objects can easily bounce off the substrates rather than stick, especially in the absence, in a vacuum, of a water layer on both surfaces to facilitate adhesion.

We describe below a reliable method for the deposition of nanoscale Au droplets originating in an ion trap in high vacuum.  Adhesion to the substrate is facilitated by melting the Au just prior to its expulsion from the trap and maintaining it in a liquid state during transit to the substrate. An electrostatic einzel lens focuses the trajectory of the particle during transit. After deposition the substrate can be removed from the vacuum via a load lock, and the solidified particles on the substrate  can be characterized by electron microscopy and elemental analysis using energy dispersive spectroscopy (EDS). 

\label{sec:expt}

\label{sec:constraints}

Practical considerations constrained our choice of projectiles and their kinetic energy at deposition.  First, the projectiles must be visible, both in the trap and on the substrate when it is imaged and measured after deposition, requiring diameter $d \geq 0.2 ~\mu$m. 
For the measurements discussed below, we use Au nanospheres with $d\cong0.25~\mu$m ($M \cong 1.6 \times 10^{-16}~$kg). After thermal treatment (necessary to remove impurities),\cite{Coppock2021} the particles typically have $Q/M \simeq 1~$C~$\perkg$ ($Q \simeq 1000~|e|$).  

Another requirement of our design is that separation between the trap and the substrate must be large enough to allow for a gate valve, which forms part of a load lock, so that the substrate may be removed and examined while the trap remains under vacuum.
This separation necessitates the introduction of an electrostatic (einzel) lens (Fig.~\ref{Apparatus}b) to localize deposition to a small region on the substrate.  Electrostatic focusing is optimized when the acceleration potential and lens potentials are large and dominate the residual patch potentials that will inevitably be present along the particle trajectory.  Since patch potentials are typically of order 1$~$V, for our experiments we have used acceleration potentials of order 100$~$V. 

The previous considerations imply an impact kinetic energy $E_k \cong1.6 \times 10^{-14}~$J (or velocity $v \cong  15~ \mpersec$). Particle adhesion can only result from inelastic collisions, but will be promoted when the surface contact adhesion energy, $E_a$, exceeds $E_k$. Values of $E_a$ for Au depend on the type of substrate and surface  preparation methods, but reported values range from 0.06 (Ref.~\onlinecite{Buks2001}) to over 1 (Ref.~\onlinecite{Hemmingson2017}) J $\permm$. Even the larger number implies that a contact area with diameter of 0.15 $\mu$m is necessary for $E_a$ to exceed $E_k$.  Thus, large deformations of the impinging particle will likely be necessary for adhesion. To make such deformations possible, we arrange for the particle to be in a liquid state at the time of impact. Liquid deformation on impact will also introduce favorable loss mechanisms to dissipate $E_k$ upon impact.

\label{sec:trap}

 In our experiments an electrically charged nanoparticle is confined in an AC quadrupole electric field trap, which consists of two metal electrodes in the shape of truncated cones, arranged coaxially, (Fig.~\ref{Apparatus}(a,b)) and is contained in a vacuum chamber.  While particles with a net positive or negative charge can be trapped, in this paper all data were taken on positively charged particles.  During normal operation, an AC voltage $\VAC$ at frequency $\nuAC$ is applied to the outer electrode of the trap and the inner electrode is grounded.  This creates a pseudopotential (Fig.~\ref{Apparatus}c) whose minimum lies about 0.7$~$mm away from the truncated edge of the outer cone along its axis of symmetry. \cite{Kane2010, Coppock2017}  In order to expel the particle from the trap for deposition, both electrodes are switched to a positive DC accelerating voltage $\Vlaunch$ during the rising part of the AC cycle (also plotted in Fig.~\ref{Apparatus}c).

\begin{figure}
\includegraphics[scale=0.35,draft=false]{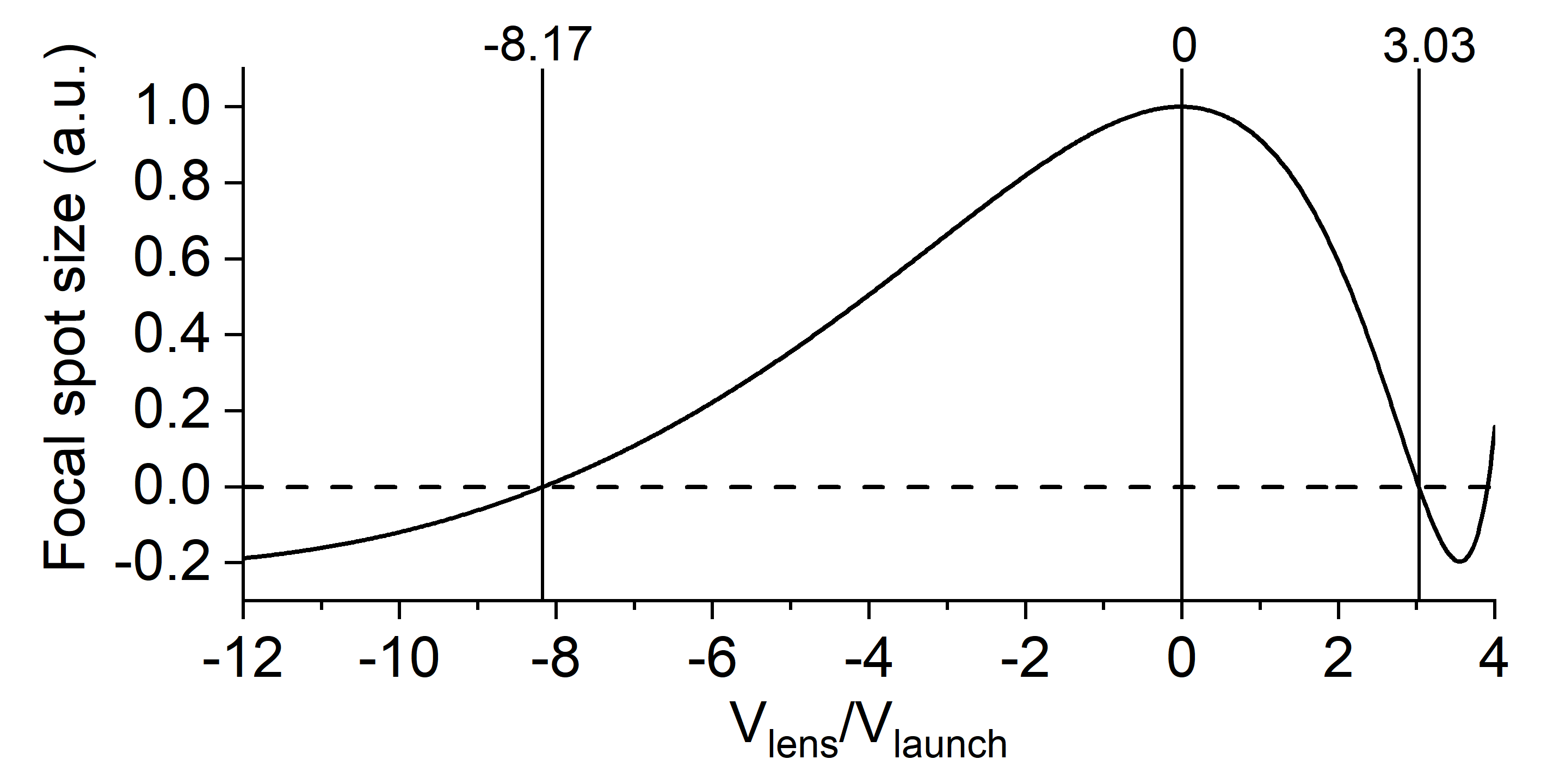} 
\caption{
Predicted size of the focused einzel lens beam at the substrate for varying ratios of lens voltage to DC accelerating voltage applied to trap at launch.  The einzel lens was modeled by first finding the axial potential (Fig.~\ref{Apparatus}d) using COMSOL Multiphysics\textregistered ~software\cite{comsol} and then using the paraxial approximation and numerical integration \cite{Hinterberger2006} to calculate the particle trajectory.  
}
\label{FocusPlot}
\end{figure}

The einzel lens (Fig.~\ref{Apparatus}b), placed between the trap and substrate, consists of metal cylinders, three grounded and two held at an electric potential $\Vlens$, arranged coaxially with the conical trap electrodes to create a two-lens architecture.  The electric potential created by the trap and lens at the moment of launch is plotted along the axis in Fig.~\ref{Apparatus}d.   If the particle starts from rest, the focal length (i.e., the distance from the point of origin at which an initially expanding beam of particles would reach a minimum magnitude spot size) depends only on the ratio of  $\Vlens$ to the accelerating voltage $\Vlaunch$.  This relationship is shown in Fig.~\ref{FocusPlot}.  The lens has multiple focal points; however, in this paper we use only negative values of $\Vlens/\Vlaunch$.  

\label{sec:detection}

The deposition substrate is a glass slide, 0.7$~$mm thick, coated on the deposition side with a transparent conductive indium tin oxide (ITO) film.  It is placed inside the vacuum chamber 236$~$mm away from the trap along its axis of symmetry (Figs.~\ref{Apparatus}b and \ref{Apparatus}e).  To facilitate removal of the substrate, a load lock is constructed by means of a gate valve, which separates the substrate from the main vacuum chamber, and a removable flange with a glass window on the opposite side of the substrate.  

Two methods are used to detect the particle at the substrate (Fig.~\ref{Apparatus}e).  A charge sensor measures the magnitude of the charge impinging on the conductive surface of the substrate, while a camera determines the location of the particle and whether it sticks.  The charge sensor is connected to the conductive coating of the substrate via a metal ring, which supports the substrate and is isolated from the other conductive parts of the apparatus by insulating posts.  It can reliably detect the impact of a particle with $Q\gtrsim800~|e|$.  The camera images the substrate from outside the vacuum chamber window with a 1:1 matched lens pair.  Its sensor area is about $5\times7~$mm and its pixel pitch is 4.4$~\mu$m.  Images are taken before and after particle deposition and compared in order to locate the particle.  Each image is averaged for 30 seconds to reduce noise.  An illuminator ($\lambda=463 \pm 25~$nm) is directed at an approximately 45 degree angle to the substrate.  A second, grounded metal ring (shown as black in Fig.~\ref{Apparatus}e), with a circular aperture of diameter 10.2$~$mm, defines the available area for deposition on the substrate.

\label{sec:temperature}

To promote adhesion, we use a laser to melt the trapped Au particle ($T_m=1337$ K) just prior to launch.  In our previous experiments,\cite{Coppock2021,Coppock2022} liquid nanoparticles were found to refreeze within 5 ms of turning off the laser.  In order to keep the nanoparticle melted during its transit to the substrate, the heating laser (Fig.~\ref{Apparatus}e) is directed along the particle’s path of flight.
A hole drilled in the central axis of the trap allows measurement of the laser power downstream of the trap.  This laser (Ushio Necsel, $\lambda=525~$nm, maximum power 5$~$W) has a broad beam (roughly 5$~$mm diameter), allowing the particle to remain illuminated during travel.  A dichroic mirror allows the camera to image along the same path as the laser travels, and a filter prevents the 525$~$nm laser light from saturating the camera.

\label{sec:procedure}

\begin{figure}
\includegraphics[scale=0.46,draft=false]{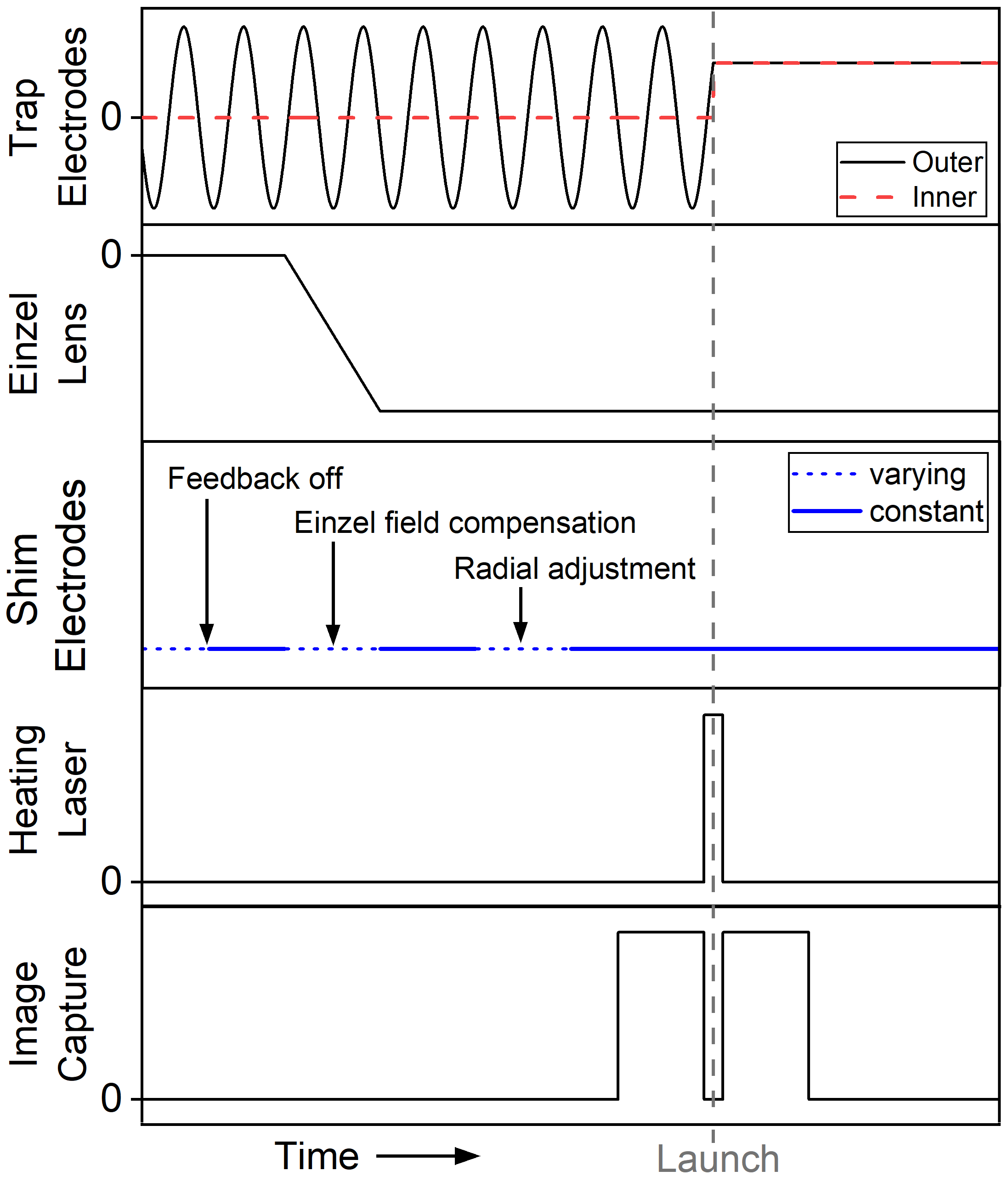} 
\caption{
Schematic depiction of the sequence of actions during the deposition process:  
trap AC frequency $\nuAC$ prior to launch is $\sim$10$~$kHz; shim and einzel electrode adjustments take about 90 s in total; the duration of each image acquisition is 30 s; and heating pulse length is typically 30--35$~$ms.  
}
\label{Timeline}
\end{figure}

Particles are prepared for deposition as follows.  Electrospray ionization is used to deliver a liquid suspension of nanoparticles to the trap and a single particle is selected for study.  The electrospray process introduces water vapor and contaminants; to minimize their effect on experiments, the particle is transferred after collection to a second trap in an adjacent vacuum chamber that is subsequently evacuated to high vacuum.\cite{Coppock2017}  In this trap, the particle undergoes a cleaning process in which it is heated with the 532 nm probe laser (Fig.~\ref{Apparatus}e) to remove any remaining contaminants from the liquid suspension and stabilize its electric charge.\cite{Coppock2021}  During this process, oxygen is added to the vacuum chamber to help remove carbon that has been observed to accumulate on the surface of the particle while it is levitated.\cite{Coppock2022}  Chamber pressure remains at approximately $1\times10^{-6}~$Torr from cleaning until deposition.  While levitated, the trapped particle is probed optically using the probe laser, and the particle's charge and mass are determined by observing discrete steps in the charge-to-mass ratio as the particle gradually discharges.  Parametric feedback cooling\cite{Nagornykh2015} is used to minimize the particle’s center-of-mass (COM) motion ($\Tcom \cong$10 K) while it is in the trap prior to expulsion.  Additional feedback is used to null the DC electric fields at the trap center using voltages applied to the shim electrodes.\cite{Coppock2021, Coppock2022}

The deposition procedure is conceptually illustrated in Fig.~\ref{Timeline}.  First, the feedback loops that cool the particle's COM motion and adjust the shim electrodes are turned off.  The feedback remains off for about 90 seconds before launch; this time is short compared to the COM thermalization time at $P\sim10^{-6}~$Torr.  Subsequent adjustments to DC potentials prior to launch are performed slowly to minimize possible heating of COM motion: the einzel lens is ramped to the desired $\Vlens$  over the course of about 60 s.  Simultaneously, a voltage is applied to all three shim electrodes to compensate for the effect of the electric field from the lens on the particle.  Next, the particle’s horizontal position is adjusted in the radial plane (normal to the trap’s axis of symmetry) using the shim electrodes.  This step ensures that launched particles impact near the center of the substrate even when $\Vlens$=0. 

When adjustments to the DC potentials are complete, a pre-deposition image is taken of the substrate.  10 ms prior to launch, the heating laser is turned on to heat and melt the trapped particle.  At launch time, the voltages on the inner and outer electrodes of the trap are simultaneously switched to a constant positive value $\Vlaunch$, pushing the particle away from the trap.  $\Vlaunch$ can be adjusted to vary the speed of impact.  Data collection for the charge sensor is also initiated at launch time. The heating laser is turned off a few milliseconds after the expected time of impact, and finally, a post-deposition image is taken of the substrate. 

\label{sec:results}

\begin{figure}
\includegraphics[scale=0.1,draft=false]{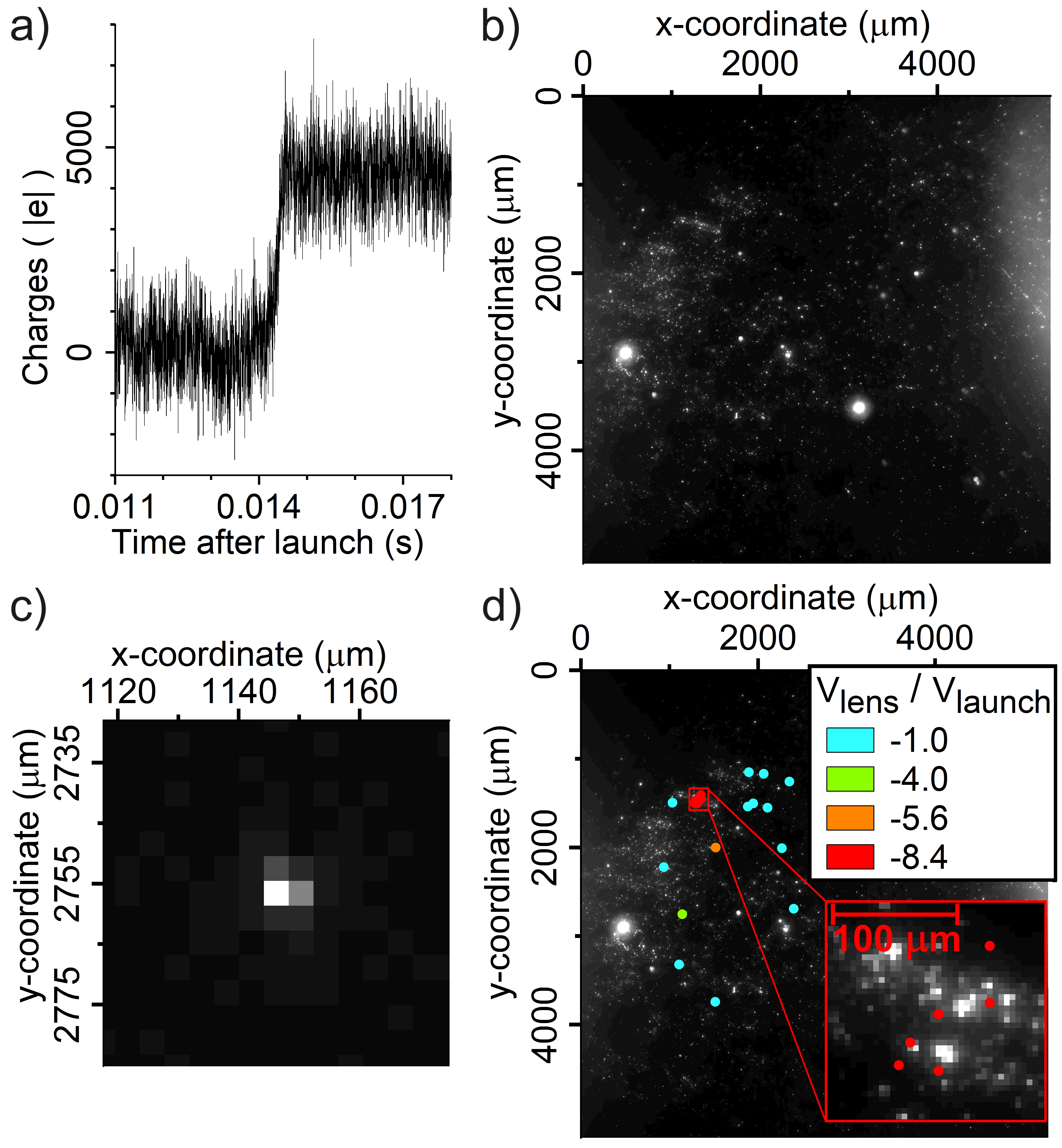} 
\caption{
(a) Charge sensor signal during particle deposition. (b) Camera image of the illuminated substrate.  Existing defects in the ITO film are numerous enough that deposited particles cannot be seen without comparing pre- and post-deposition images.  (c) Subtracted post-deposition minus pre-deposition image, cropped to show (14x14)-pixel area near the deposited particle.   (d) Map of twenty deposited particle positions on the substrate.  Six particles launched at a close-to-optimal focus (red dots) landed in an area of diameter 120$~\mu$m.
}
\label{ChargeCamera}
\end{figure}

Typical charge sensor data are shown in Fig.~\ref{ChargeCamera}a. 
An image of the substrate is shown in Fig.~\ref{ChargeCamera}b.  The substrates contain numerous small film and surface defects with sizes ranging from 100 nm to 10 $\mu$m, while a newly deposited nanoparticle adhering to the slide causes an increase in brightness in only one or a few pixels of the image. Consequently, optical detection of the adhered particle requires the comparison of pre- and post-deposition images.  

In initial tests, the projectiles were not melted and had impact velocities in the range of 40--70$~\mpersec$.  While a charge signal was observable, no particle was imaged on the substrate, indicating that they were not sticking. We reduced the impact velocity as low as 8$~\mpersec$ in an attempt to make $E_k$ smaller than $E_a$, but still no particles were visible in the camera image.  To solve this problem, the particles were melted during deposition using the heating laser (see Figs.~\ref{Apparatus}e and \ref{Timeline}).  For the data presented here, the intensity of the laser near the center of the beam was approximately $2\times10^{4}~\Wpermm$, which heats a gold particle of 250$~$nm diameter above 1500$~$K. 

Once the melting procedure was incorporated, deposited particles were easily found in the camera images using an automated analysis program that located areas with large differences between the post- and pre-deposition images. Deposition was confirmed (i.e., particles were detected with both the charge sensor and the camera) in 19 out of 20 attempts.
As a final adjustment, the quantity $\Vlens/\Vlaunch$ was varied in order to improve the lens focus.  A setting of $\Vlens/\Vlaunch=-8.4$ caused a series of six particles to fall within a circle of diameter 120$~\mu$m (Fig.~\ref{ChargeCamera}d).
For these launches, the amplitude of $\VAC$ varied from 300 to 400$~$V, and $\Vlaunch$ varied from $15\%$--$60\%$ of the AC voltage.  Estimated impact velocities ranged from 17 to 26$~\mpersec$.

\begin{figure}
\includegraphics[scale=0.26,draft=false]{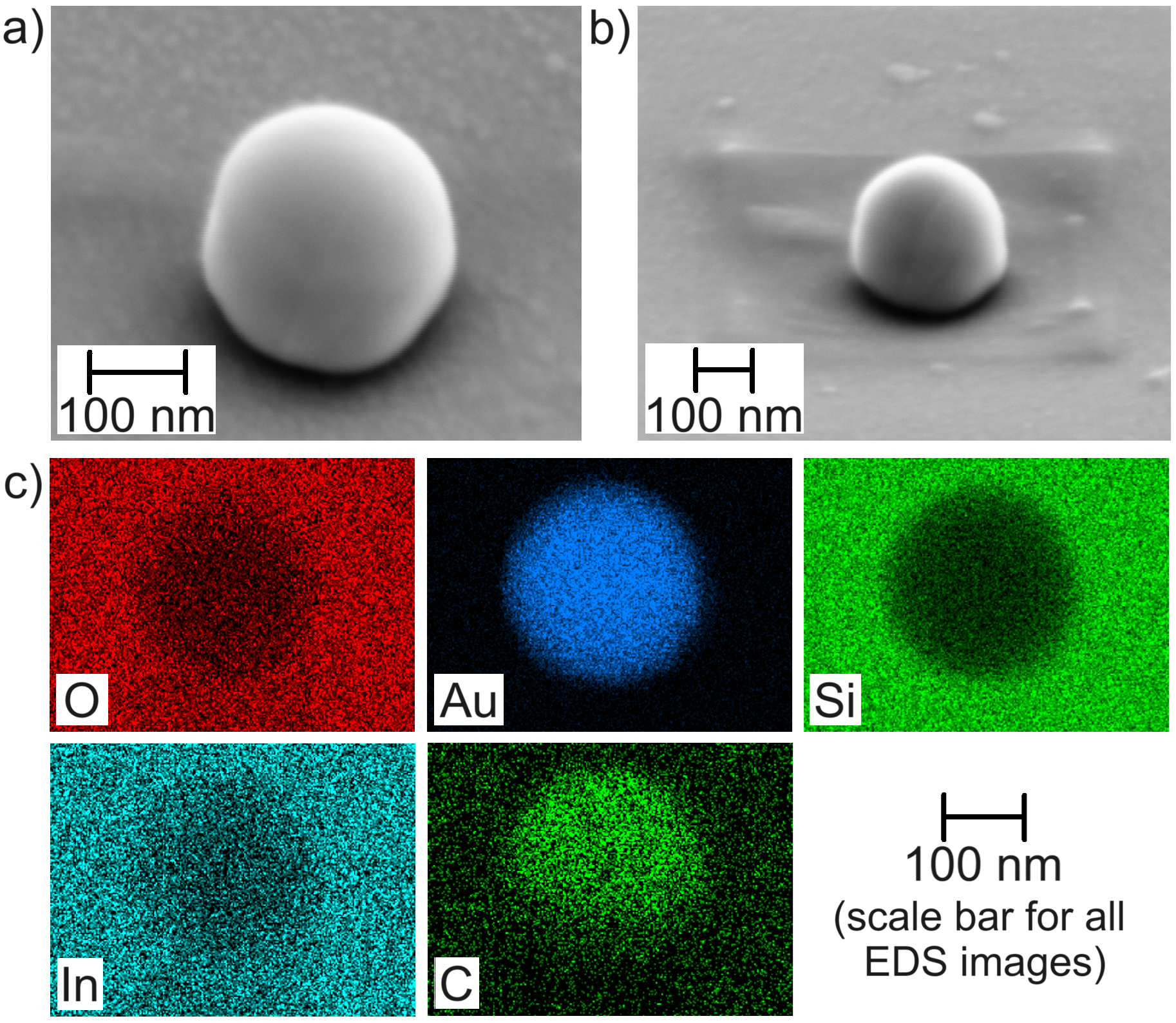} 
\caption{
SEM images of a single Au particle deposited on a glass substrate with an indium tin oxide coating.  The particle diameter (inferred from the mass measured while the particle was levitated, just prior to deposition) was 257$~$nm.
(a) SEM image with detector at a 35\degree~angle from normal with respect to the substrate.  (b) SEM image with detector at a 50\degree~angle from normal.  (c) Energy dispersive spectroscopy maps showing the spatial distribution of the five most abundant elements in an image of the same deposited Au particle.  Brightly colored pixels indicate the presence of the labeled element, while black pixels indicate its absence.
}
\label{SEM}
\end{figure}

After deposition, the substrate was removed from the vacuum chamber and examined using a scanning electron microscope (SEM) with energy dispersive spectroscopy (EDS) capability. Three particles were identified at locations corresponding to those in the substrate camera images (Fig.~\ref{ChargeCamera}d) and their size and composition were confirmed.  Data for one of these particles are shown in Fig.~\ref{SEM}.  SEM images in Figs.~\ref{SEM}a and \ref{SEM}b show a particle of size similar to that measured just before launch.  EDS maps in Fig.~\ref{SEM}c show Au localized inside the diameter of the particle.  The only other element that is enhanced inside the diameter of the particle is C, providing further evidence for our previous observations \cite{Coppock2022} of carbon incorporation in levitated Au.  

\label{sec:conclusions}

In conclusion, we have demonstrated reliable deposition of levitated nanoscale material, in the form of liquid Au droplets, and shown that projectile droplets can be focused to an area of diameter 120$~\mu$m 236$~$mm away from the launch point.  Finer focusing has been impracticable due to the low ($\leq$ 1 /day) throughput of our current sample preparation and deposition process; however, further work to improve the focus could be rewarding.  The center-of-mass motion of nanoscale particles in similar systems\cite{DoubleDania} has been cooled to $\Tcom \cong 10~$mK.  A trapped particle at this temperature in our apparatus ($M=1.6 \times 10^{-16}~$kg, $\nuAC=1~$kHz)  would be localized to $\sim$5$~$nm. While the simple rotationally symmetric einzel lens will have inevitable aberrations, the slow speed of particles moving through the lens means that dynamic corrections could be applied to the lens voltages for aberration correction.\cite{Schonhense2002,Jacob2016}  It is thus possible that a future system could deposit particles onto a substrate with nm accuracy. 

Our technique should be applicable to a variety of liquids and potentially soft solid projectiles. Liquid Au projectiles could also be useful as ``tracers" when more challenging materials (either harder to see or less likely to adhere) are being deposited.  Reliable deposition of arbitrary solid projectiles will either require orientation of solid surfaces to the substrate to improve adhesion or the use of slower projectiles than we have used so that there is less kinetic energy to dissipate and adhesion is more likely.

Our ability to control the temperature and velocity of the nanoparticle during deposition opens up possibilities for future studies of impacts of liquid droplets.  The deposited particles in our experiments (Fig.~\ref{SEM}) appear slightly non-spherical.  SEM or other imaging techniques could be used to precisely measure the deformation of the particles upon impact.

\begin{acknowledgments}
This work was supported by the Laboratory for Physical Sciences.
\end{acknowledgments}

\section*{Conflict of Interest}
The authors have no conflicts of interest to disclose.
\section*{Author Contributions}
\textbf{Joyce E. Coppock:} Investigation (lead); Methodology (equal); Writing -- original draft (equal); Writing -- review \& editing (equal).  \textbf{B. E. Kane:} Conceptualization (lead); Funding acquisition (lead); Writing -- original draft (equal); Writing -- review \& editing (equal).
\section*{Data Availability}
The data that support the findings of this study are available from the corresponding author upon reasonable request.

\section*{References}

\bibliography{deposition}

\end{document}